\documentclass[article]{aa}
\usepackage{graphicx}

\newcommand{\prp}    {${\rlap.}^{\prime}$}
\newcommand{\grp}    {${\rlap.}^{\circ}$}
\newcommand{\pri}    {${\rlap.}^{\prime \prime}$}
\newcommand{\rl}     {${\rlap.}^{s}$}

\newcommand{\ltsima} {$\; \buildrel < \over \sim \;$}
\newcommand{\simlt}  {\lower.5ex\hbox{\ltsima}}            
\newcommand{\gtsima} {$\; \buildrel > \over \sim \;$}
\newcommand{\simgt}  {\lower.5ex\hbox{\gtsima}}            

\begin{document}

\title{Faint arc-minute extended radio emission around Cygnus X-3}

\author{
J.~R. S\'anchez-Sutil\inst{1}, J. Mart\'{\i}\inst{2,1}, J.~A. Combi\inst{2,1}, P. Luque-Escamilla\inst{3,1}, A.~J. Mu\~noz-Arjonilla\inst{2,1}, J.~M. Paredes\inst{4} and G. Pooley\inst{5}
}

\offprints{J. Mart\'{\i}}

\institute{Grupo de Investigaci\'on FQM-322,
Universidad de Ja\'en, Campus Las Lagunillas s/n, Edif. A3, 23071 Ja\'en, Spain \\
\email{jrssutil@hotmail.com}
\and Departamento de F\'{\i}sica, EPS,  
Universidad de Ja\'en, Campus Las Lagunillas s/n, Edif. A3, 23071 Ja\'en, Spain \\
\email{jmarti@ujaen.es, jcombi@ujaen.es, ajmunoz@ujaen.es}
\and Dpto. de Ing. Mec\'anica y Minera, EPS,
Universidad de Ja\'en, Campus Las Lagunillas s/n, Edif. A3, 23071 Ja\'en, Spain \\
\email{peter@ujaen.es}
\and Departament d'Astronomia i Meteorologia, 
Universitat de Barcelona, Mart\'{\i} i Franqu\`es, 1, 08028 Barcelona, Spain \\
\email{jmparedes@ub.edu}
\and Astrophysics, Cavendish Laboratory, J. J. Thomson Avenue, Cambridge CB3 0HE, UK\\
\email{guy@mrao.cam.ac.uk}
}

\date{Received / Accepted}

\authorrunning{S\'anchez-Sutil et al.}
\titlerunning{Faint arc-min extended radio emission around Cyg X-3}

\abstract
{}
{We revisit the vicinity of the microquasar Cygnus X-3
at radio wavelengths. We aim to improve our previous search for possible associated extended radio features/hot spots 
in the position angle of the Cygnus X-3 relativistic jets focusing
on shorter angular scales than previously explored.
}
{Our work is mostly based on analyzing modern survey and archive radio data, mainly including observations
carried out with the Very Large Array and the Ryle Telescopes. We also used deep 
near-infrared images that we obtained in 2005.
}
{We present new radio maps of the Cygnus X-3 field computed after combining multi-configuration 
Very Large Array archive data at 6 cm and different observing runs at 2 cm with the Ryle
Telescope. These are probably among the deepest radio images
of Cygnus X-3 reported to date at cm wavelengths.
Both interferometers reveal an extended radio feature within a few arc-minutes of the microquasar position,
thus making our detection more credible. 
Moreover, this extended emission is possibly non-thermal, although this point still needs confirmation. 
Its physical connection with the microquasar is tentatively considered
under different physical scenarios. We also report on the serendipitous discovery of a likely
Fanaroff-Riley type II radio galaxy only $3^{\prime}$ away from Cygnus X-3.
 }
{}

\maketitle

\section{Introduction}

Cygnus X-3 is one of the X-ray binaries considered as the prototype of the family of
galactic sources with relativistic jets, i.e., microquasars. It was originally discovered in X-rays
by \cite{gia67}. This system made headlines more than three decades
ago when its first strong radio outbursts were detected, as described in \cite{g72} and subsequent papers.
Such flaring events have been repeating since then, typically one or two times per year, 
with flux density increments of almost two orders of magnitude above the normal quiescent
level of $\sim0.1$ Jy at cm wavelengths (\cite{w1994}). At present, Cygnus X-3 is considered to be  
a high-mass X-ray binary (HMXB) with a WN Wolf-Rayet (WR) companion star of WN8 type (see e.g. \cite{ke1996} and \cite{ko2002}).
The nature of the compact object is not well constrained, and 
neither a neutron star nor a black hole accreting from the WR star can be ruled out
(see e.g. \cite{er1998} and \cite{mi1998}). 
The observed X-ray (\cite{p1972}) and infrared (\cite{bk1973})
modulation of 4.8 h is believed to be connected with the system orbital period, which
is rather short for an HMXB. Orbital-phase-resolved spectra of Cygnus X-3 in the near infrared are consistent
with it (\cite{han2000}).
The distance has been estimated to be approximately 9 kpc (\cite{pred00}), which  
agrees with a strong interstellar absorption ($A_V \sim 20$ mag, \cite{ke1996}) that renders the optical counterpart 
undetectable in the visual domain.

\begin{table*}
\begin{center}
\caption[]{\label{vlaobs} VLA archive observations used in this paper \label{obslog}}
\begin{tabular}{cccccccc}
\hline
\hline
Date  &  wavelength  & VLA  & Number of  & Bandwidth  & Number of    &  Project & Time on    \\
      &  (cm)   & configuration. &  IFs  &  (MHz)      &  visibilities &  id. & source (hrs) \\ 
\hline 
1992 Jun 8 &   6 &  DnC & 2 & 50 & 432229  & UT002 &  5   \\
1997 May 4 &   6 &  B & 2  & 50 & 718939 & AM551   &  5.7   \\ 
2000 Aug 29 &  6 &  D & 2 & 50 & 29167 & AH669     &  0.33    \\
\hline
\hline \end{tabular}
\end{center}
\end{table*}

Galactic microquasars such as Cygnus X-3 are known to release a significant amount of energy
in the form of collimated relativistic jets into their surrounding inter-stellar medium (ISM).
For Cygnus X-3, the averaged energy injection rate of its relativistic jets is estimated to be
at least $10^{37}$ erg s$^{-1}$ (\cite{m2005}), 
i.e., enough to supply $\sim 10^{50}$ erg (10\% of a supernova explosion)
during the expected lifetime of the WR star.
The Cygnus X-3 jets have been repeatedly resolved  
as sub-parsec transient radio features, propagating at a significant fraction of
the speed of light in the north-south direction, thanks to interferometric radio techniques 
from arc-second (\cite{m2001}) to milli arc-second angular scales (e.g. \cite{g1983,miller04,tu2007a}).
However, up to now there is no robust evidence of interaction between the relativistic ejecta
of the system and the surrounding ISM on larger, few pc scales. This is in contrast with other 
microquasars, such as Cygnus X-1 or Circinus X-1, where a clear signature of interaction between 
their relativistic jets
and the ISM does exist  
with a ring-like or lobe morphology 
several pc wide (\cite{ga2005,tu2007b}).
Additional examples of relativistic jet/ISM interaction come
from the also pc-scale lobes of the Galactic Centre Annihilator 1E1740.7$-$2942 (\cite{mira92})
or GRS 1758$-$258 (\cite{m2002}). The detection of such ring/lobe features is a useful tool
to better constrain the system's energetics. 


The search for large scales features associated with the Cygnus X-3 radio jets has been a concern of the
authors during recent years. Two H\ion{II}\ regions located $\sim40^{\prime}$ from Cygnus X-3 
along the jet position angle were first tentatively proposed as possible large scale lobes (\cite{m2000}).
However, no further evidence for a physical connection could be found beyond the mere geometric alignment. 
A closer search revealed later the existence of two possible hot spot candidates (HSCs) associated with Cygnus X-3, 
thus suggesting an analogy with Fanaroff-Riley type II (FR II) radio galaxies (\cite{m2005}). 
The apparent hot spots were two faint radio sources with non-thermal spectra
at angular distances of 7\prp 07 and 4\prp 36 from Cygnus X-3. The line joining them was also within one degre
of the almost North-South position angle of the inner arc-second radio
jets (\cite{m2001}). Unfortunately, follow up radio and near infrared observations 
of both the HSCs and the Cygnus X-3 nearby environment did not confirm the proposed 
hot spot nature and indicated that they were most likely background or foreground objects (\cite{m2006}). 
This fact left open again for Cygnus X-3 the issue of searching for signatures of energy 
deposition from its relativistic jets into the ISM.

In this context, the main purpose of this paper is to
present new very deep radio images of Cygnus X-3
and its environment obtained after combining multi-epoch archive and survey data,
together with our own observations.
The resulting maps enable us to put very strong limits for any extended or compact radio features
that could be associated to Cygnus X-3 on a scale of a few pc. 
We also report on
several extended radio features in the field with apparent non-thermal spectra that were previously unknown.
Their possible connection with the microquasar is discussed from a skeptical point of view.


\section{Deep VLA radio map of the Cygnus X-3 vicinity}

\subsection{Archive VLA 6 cm data analyzed in this paper}

Most observations used in this paper come from archive data obtained with the
Very Large Array (VLA) operated by the National Radio Astronomy Observatory (NRAO) in the USA.
By similarity with the Cygnus X-1 or Circinus X-1 cases, the possible large scales radio features around Cygnus X-3
(if any) are likely to be a few pc extended and, therefore, with few arc-minute angular sizes.
The more compact D configuration of the VLA appears thus
as the best choice for our detection purposes at these angular scales.
Concerning wavelength, we decided to use 6 cm data which provides a Full-Width Half Maximum (FWHM) 
primary beam of about $9^{\prime}$. Then, a 1 to 10 pc feature 
would be well covered within the more sensitive, 
inner part of the primary beam.

Surprisingly, the NRAO data archive contains no VLA observations for Cygnus X-3 in 
the pure D configuration of the array and with more than a half hour of integration time.
The archive data more closely matching our requirements comes from the hybrid DnC configuration.
We also included one B configuration run for enhanced angular resolution in addition to sensitivity
on arc-minute scales. The log of the observing runs that we have combined into a single radio image
is given in Table \ref{vlaobs}.




\begin{figure}[htpb]
\begin{center}
\vspace{8.5cm}

\includegraphics{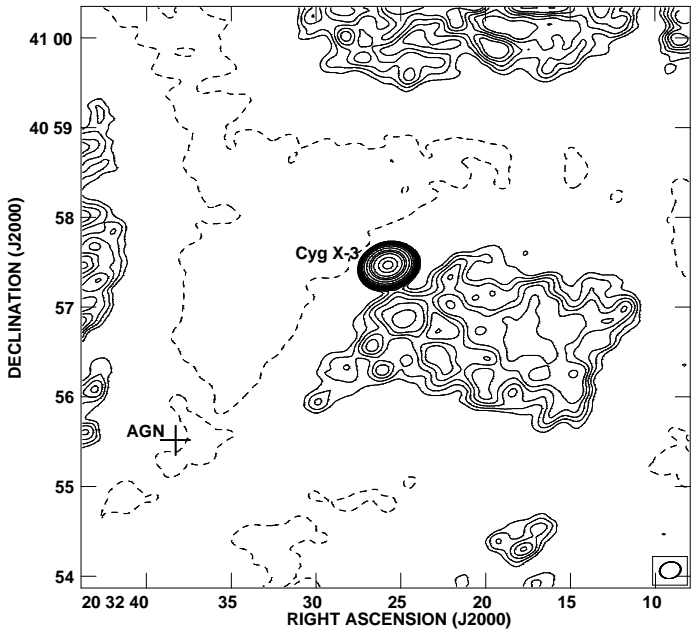}

\caption{VLA map of Cygnus X-3 at the 6 cm wavelength using visibility data in the DnC and D configuration
of the array. The central variable core of the microquasar has been subtracted as described in the text
and replaced by a constant point source with an average flux density of 100 mJy.
Extended radio emission in the vicinity of Cygnus X-3 is clearly seen in this map.
Contours shown correspond
to $-4$, 4, 5, 6, 7, 8, 10, 12, 15, 20, 30, 50, 100, 200, 300, 500, 1000 and 1500 times 57 $\mu$Jy beam$^{-1}$,
the thermal rms noise of the map. The negatives in this low-resolution map go significantly deeper ($-10\sigma$) than
shown for clarity purposes of the figure. This occurs only in some pixel areas inside the largest dashed contour.
The synthesized beam is shown as an ellipse in the bottom right corner. 
It corresponds
to 14\pri 17 $\times$ 10\pri 95 with a position angle of $-$75\grp 8. The cross marks the core position of an AGN also reported in this work.
}
\label{cd+d} 
\end{center} 
\end{figure}

The data were processed using the AIPS software package of NRAO following the standard procedures for continuum calibration 
of interferometers. Moreover, the {\it u-v} data were self-calibrated in phase. 
Due to the significant flaring variability of the Cygnus X-3 core during the observations the standard CLEANing methods could not be applied directly. 
Instead, we proceeded following a method similar to that outlined in Mart\'{\i} et al. (2000). 
Therefore, we had to subtract a time-variable point source from the visibilities in the {\it u-v} plane
using different AIPS tasks. First of all, we split the data in blocks of less than 5\% amplitude variation. 
For each block, we derived the flux density of the component to be subtracted at the microquasar core position
with task UVFIT. Afterwards, we removed it using UVSUB and 
all almost variability free blocks were finally recombined using DBCON.

The next step was to combine the data of the three projects we are handling. 
At this point, the UVFIX task of AIPS had to be used to set a common phase center 
to allow the appropriate combination of the visibilities. Finally,
the DBCON task was used again to merge all data sets in Table \ref{obslog} into a single $uv$ file.
This was performed with an appropriate weighting to enhance the short baselines of DnC+D configuration that
give sensitivity to extended emission.
A constant point source, with the average flux density of Cygnus X-3, was finally added at the position of the subtracted core.

\subsection{Extended emission}

We first mapped our multi-configuration data without using the long baseline visibilites provided
by the B configuration of the array. Fig. \ref{cd+d} shows a contour plot of the field obtained
using the IMAGR task of AIPS with the ROBUST parameter set to five and DnC+D data only.
The primary beam correction has been applied
to the radio map restricted to its 50\% contour.
Such a conservative choice was adopted to avoid any problem towards the otherwise noisy edges of the map.
Remarkably, Fig. \ref{cd+d} shows that Cygnus X-3 is located just at the edge of an extended 
radio feature clearly visible towards the South and South-West direction.
Despite some deep negative artifacts, likely due to missing short baselines and perhaps to the subtraction
procedure as well, the reality of the extended emission is beyond
doubt thanks to confirmation by independent radio instruments (see below).  
In addition, two clumps of extended emission are also present to the North and East from the microquasar, 
respectively, but well separated by several arc-minutes.

Despite the morphological suggestion from Fig. \ref{cd+d}, it is not possible
to conclude whether or not the extended emission is physically associated to Cygnus X-3. 
While unrelated extended emission from the Galactic Plane cannot be ruled out,
we could also be seeing traces of a disturbed lobe with an emission plume extending 
sideways for as long as 5 pc at the Cygnus X-3 distance. 
Similar scaled-up effects are very often seen in the large scale lobes and plumes of radio galaxies, 
where the interaction of the jet pointing towards the observer
with the intergalactic medium induces turns in the flow direction changing it by as much as $90^{\circ}$. 
A representative example of this phenomenon is found, for instance, in the large lobes of 3C 264 (\cite{l2004}) 
where the most remarkable feature is the transition between a well collimated narrow jet at distances 
from the core below 80 pc, to a conical-shaped wide jet, with a large opening angle. 
The possibility than an analog physical process could take place in Cygnus X-3
is a scenario worth exploring based on this evidence. Moreover,
as in 3C 264, we also find in our map a more distant patch of extended emission 
right at the expected position angle of the counterjet.

\begin{figure*}[htpb]
\begin{center}
\vspace{16.5cm}

\includegraphics{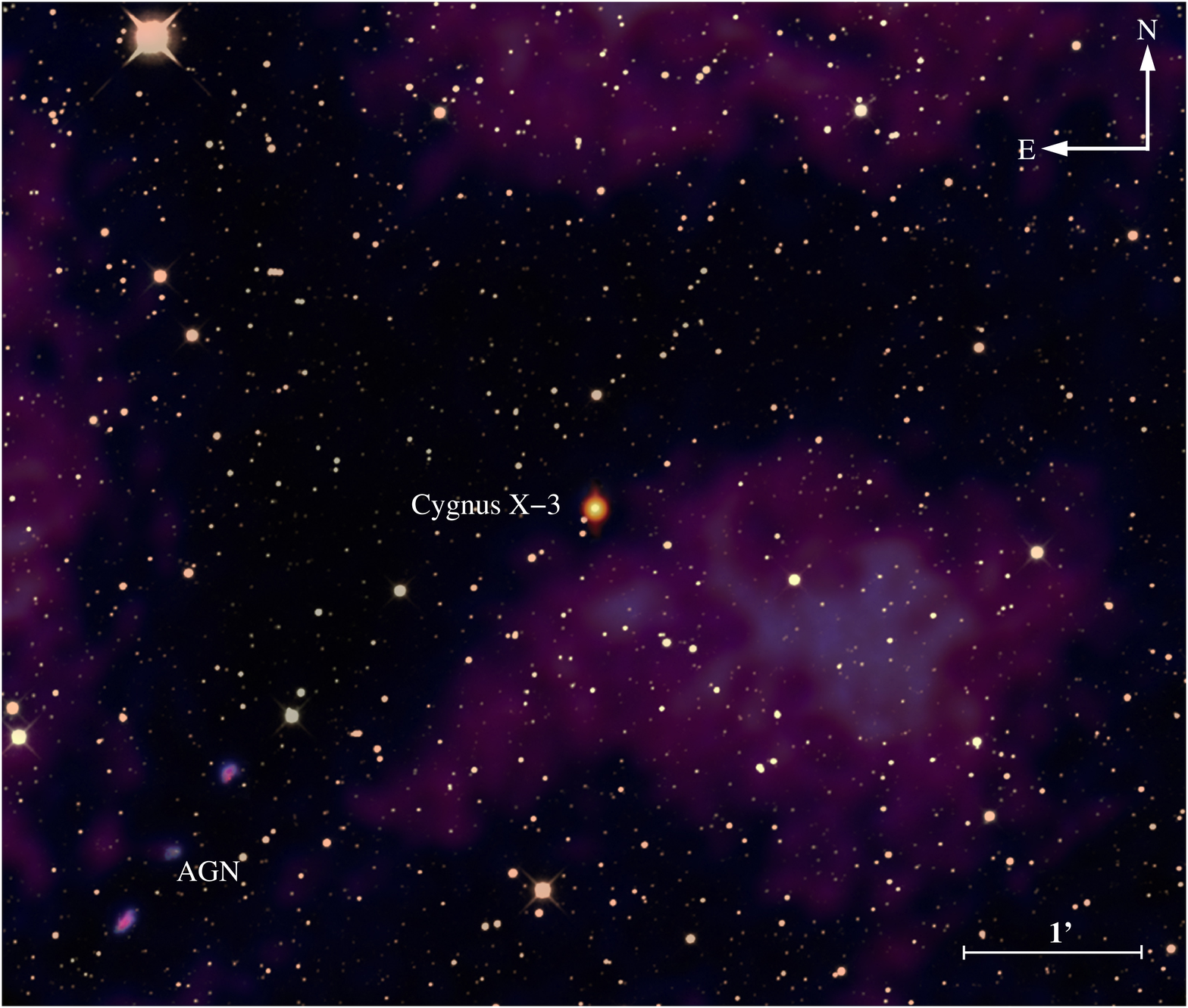}

\caption{Composite radio and near infrared image centered on Cygnus X-3
and covering a $5.6 \times 6.7$ arcmin$^2$ field of view. 
The bluish layer of this montage corresponds to our
VLA map at the 6 cm wavelength computed by combining visibilities
from the VLA projects UT002 (DnC configuration), AH669 (D-configuration) and AM551 (B-configuration).
The background layer overlays a $Ks$-band near infrared image of the same field (\cite{m2006}).
Traces of the Cygnus X-3 arcsec radio jets are also visible.
These come from the AM551 data obtained weeks after a giant radio outburst (\cite{m2000}). In addition a triple radio source
is clearly seen at the bottom left corner which is almost certainly a background AGN.
We interpret it as a FR II radio galaxy with its likely core and two hot spot components perfectly aligned.
}
\label{cd+d+b}
\end{center}
\end{figure*}

The resulting map obtained after combining the full DnC+D+B configuration data at 6 cm
is presented in Fig. \ref{cd+d+b}. It is
part of a composite image containing both radio (this work) and near infrared data (\cite{m2006}).
This montage is included 
to render a global and enhanced view of the different source components 
present in the complex Cygnus X-3 field (e.g. extended emission, compact cores, galactic and extragalactic jets, etc.)
which otherwise would be difficult to visualize simultaneously.
As quoted above, the Cygnus X-3 variable radio core was removed to avoid variability artifacts
during the CLEANing process. A compact point source, with the average
flux density level, was later added later at its location for illustration purposes. At this point,
The synthesized beam was practically a circular Gaussian with 1\pri 5 angular size
and the map rms noise 9.5 $\mu$Jy beam$^{-1}$. Correction for primary beam response was also applied to the radio data
up to a conservative response value of 50\% for appropriate physical measurements. As a final cosmetic step,
the extended radio emission next to Cygnus X-3 and whose nature is still unclear
has been slightly smoothed for better display.

\subsection{A triple radio source in the Cygnus X-3 field} \label{triplesec}

\begin{figure}[htpb] 
\begin{center} 
\vspace{8.5cm} 
\includegraphics{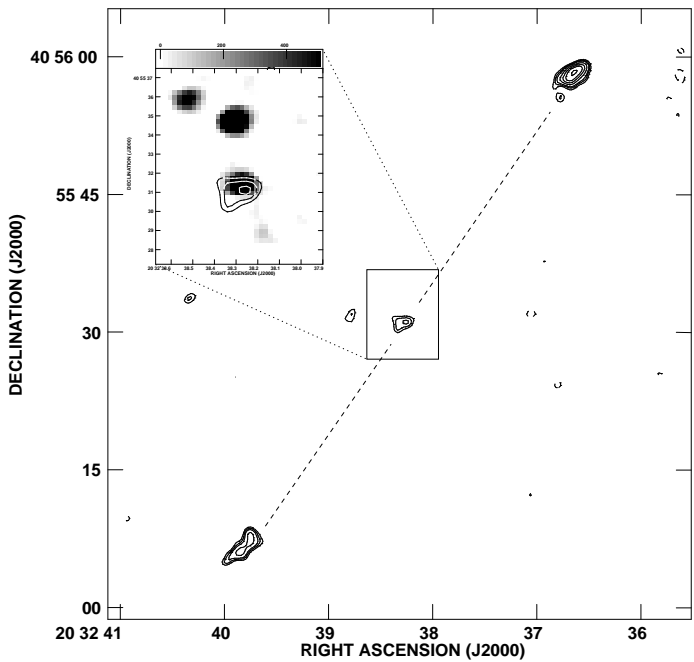}         
\caption{Detailed view of Fig. \ref{cd+d+b} showing a contour map of
the triple radio source discovered in the close vicinity of Cygnus X-3 at the 6 cm wavelength
and believed to be a background AGN.
This object clearly consists of a central core with two symetrical and well aligned hot spot
components and is likely an FRII radio galaxy. Radio contours shown correspond to
$-4$, 4, 5, 6, 8, 10, 12 and 15 times 9.5 $\mu$Jy beam$^{-1}$. 
The panel inset shows a zoom of the central component with the 
background grey scale representing the $Ks$-band image from Calar Alto observatory (\cite{m2006}).
A stellar-like near infrared counterpart candidate is obvious.
}
\label{triple}
\end{center}
\end{figure}

The radio map in Fig. \ref{cd+d+b} is also deep enough to reveal a 
a triple radio source with a compact core and two aligned lobe/hot spots components previously unknown.
The location of the proposed core is at coordinates  
$\alpha_{\rm J2000.0}= 20^h 32^m $38\rl 29$\pm$0\rl 02 and
$\delta_{\rm J2000.0}= +40^{\circ} 55^{\prime}$31\pri 0$\pm$0\pri 2,
with a 6 cm flux density of $0.27 \pm 0.06$ mJy.
Its angular distance from Cygnus X-3 is 3\prp 06.
The overall morphology of this triple source consistent with a FR II radio galaxy, as shown in Fig. \ref{triple}. 
The grey scale in Fig. \ref{triple} comes from previous near infrared $Ks$-band observations
(\cite{m2006}), where a point-like counterpart with magnitude $Ks=17.2 \pm 0.1$ is coincident
with the compact core within astrometric error. Microquasars and AGNs are often considered
as close relatives with the same accretion/ejection processes going on (\cite{mr1999}).
To our knowledge, the Fig. \ref{cd+d+b} map
is the first case where representatives of both families can be simultaneously imaged and resolved
within the same primary beam thus giving it a strong educational value. 
This fact also opens a possibility for proper motion studies of Cygnus X-3
using future very sensitive interferometers with respect to the FR II core.

\section{An independent detection of extended emission in the Cygnus X-3 vicinity}

\begin{figure}[htpb]
\begin{center}
\vspace{8.5cm}

\includegraphics{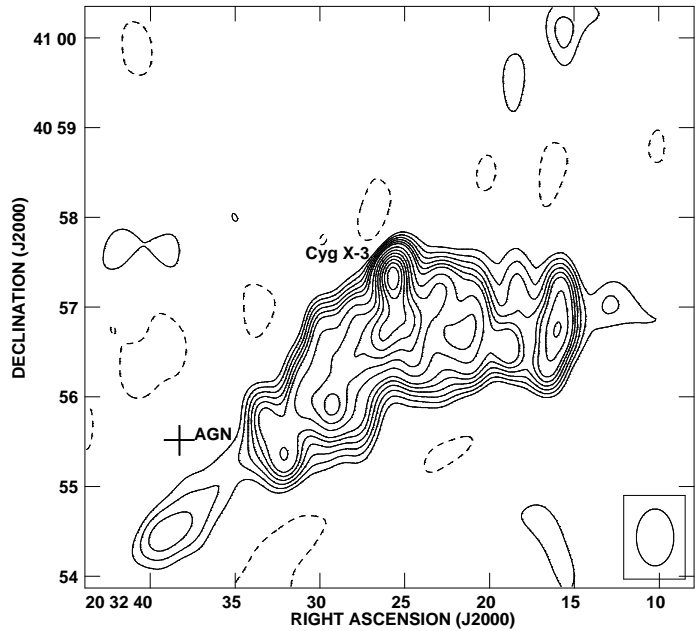}

\caption{Cygnus X-3 as observed in its `quenched' state with the Ryle Telescope  at the 2 cm wavelength.
Contours shown correspond to $-3$, 3, 4, 5, 6, 7, 8, 10, 12, 14, 16, 18, 20 and 22 times the
0.15 mJy beam$^{-1}$, the rms thermal noise. The synthesized beam is shown at the bottom
right corner and corresponds to an ellipse of 
$38^{\prime\prime} \times 25^{\prime\prime}$, with position angle of $0^{\circ}$.
This map is uncorrected for primary beam response (FWHM $6^{\prime}$, centred on Cygnus X-3) 
and covers exactly the same field
as Fig. \ref{cd+d}. Both the position of Cygnus X-3 and the nearby AGN in the field are indicated.} 
\label{mrao}
\end{center}
\end{figure}

The Ryle Telescope (Cambridge) has for some years been used in a
`spare-time' mode for monitoring of X-ray binaries; for details,
see \cite{pf97}. In recent years it has operated
at 15\,GHz ($\lambda = 2\,\rm{cm}$).
Cygnus\,X-3 is unique in that it can show
periods of very low (`quenched') flux density for some days
immediately before major radio/X-ray flares (\cite{w1994}).
This fact is potentially useful in giving warning
of the flares.
Observations during this phase had demonstrated
that there is some emission
on larger angular scales which had so far been ignored or
treated as a complication in the analysis.
Essentially this emission is readily detected only on the
shortest baseline of the telescope (about $1800 \lambda$),
for only about two hours of the hour angle range.
The apparent flux density at 15\,GHz of this extended
structure is about 35\,mJy.
The report of the VLA maps shown in this paper prompted a
reanalysis of some recent data: we used 8 observations
during the period 2006 Jan 20 -- 29  (excluding Jan 26,
when Cygnus\,X-3  reached 50 mJy). The mean flux density of this
HMXB for the remaining days was about 3 mJy.
The data, covering time-spans from about 1 h up to 12 h,
were mapped and the resulting data cleaned.
The synthesised resolution is $25^{\prime\prime} \times 38^{\prime\prime}$.
No attempt was made at any self-calibration or allowance
for the variable nature of the core source.
Neither has the map been corrected for the primary beam
of the telescope.

The resulting image is shown in Fig. \ref{mrao}.
This must be interpreted with some caution, since
(a) the Fourier plane is clearly not well-sampled;
(b) the structure extends over most of the primary beam
of the telescope ($6^{\prime}$ FWHM). 

Despite the frequency difference,
the agreement of the overall structure mapped by two
independent interferometers in Figs. \ref{cd+d} and \ref{mrao} gives robustness to the detection of arc-minute
extended emission close to Cygnus X-3.  
The shape of the 2 cm emission is here more suggestive of a physical North-South elongated
connection with Cygnus X-3 than at 6 cm. Moreover, the extended emission  
also appears reminiscent of an arc-like morphology opening towards the South and
with the microquasar in the middle. Could we be seeing part of a ring-like structure
similar to that detected around Cygnus X-1 by \cite{ga2005}? Having in mind all image limitations quoted above,
it is currently  difficult to discriminate among
a lobe, ring or a simple unrelated emission scenario until better observations are available.

\section{Spectral index map of the extended emission}

\begin{figure*}[htpb]
\begin{center}
\vspace{8.3cm}
\includegraphics{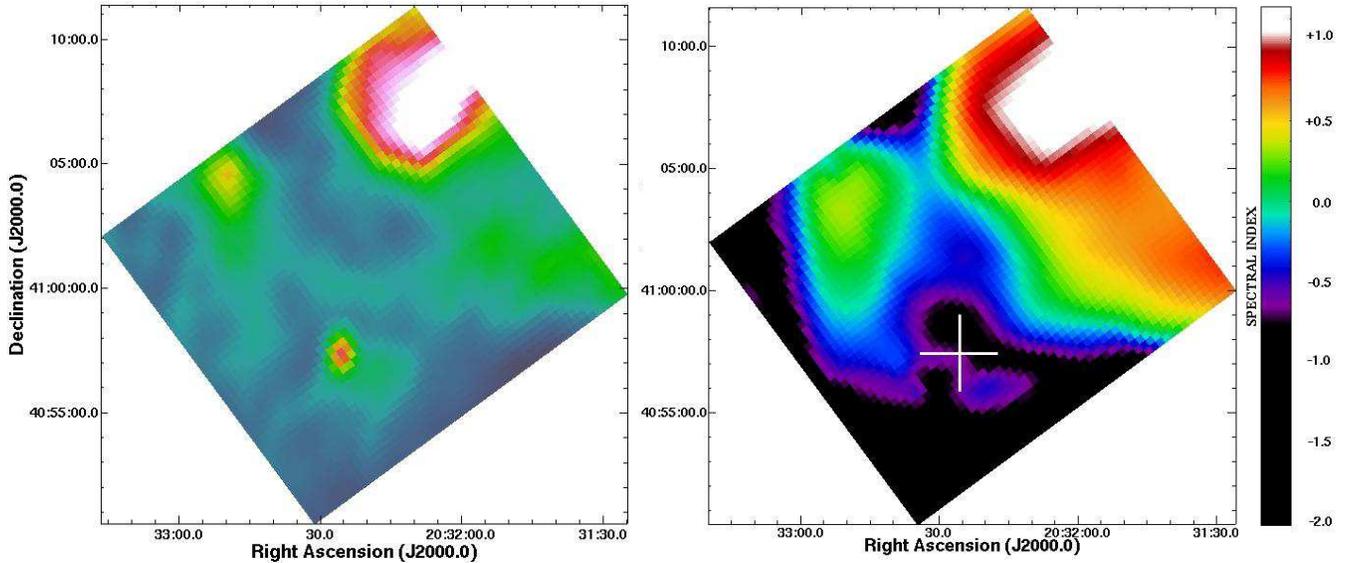}
        \caption{{\bf Left.} Map of Cygnus X-3 at 1.4 GHz obtained from the Canadian Galactic Plane Survey (CGPS). This map shows some traces of the extended emission discovered around Cygnus X-3 in the VLA and Ryle Telescope maps,
thus reinforcing its reality. {\bf Right.} Spectral index map of Cygnus X-3 computed using images from the Canadian Galactic Plane Survey (CGPS) at 1.4 GHz and the Green Bank 4.85 GHz northern sky survey (GB6). The map shows hints of non-thermal emission
mainly coincident with the southern extended emission discovered around Cygnus X-3 in the VLA map.
The big cross indicates the instrumental FWHM of the Cygnus X-3 central core subtracted when creating
this spectral index map.}
        \label{surveys}
\end{center}
\end{figure*}

We also inspected the Canadian Galactic Plane Survey (CGPS, \cite{tay03})
at 1.4 GHz and the Green Bank 4.85 GHz northern sky survey (GB6, Condon et al. 1991, 1993, 1994). 
Some traces of the extended feature quoted in previous sections are also visible around Cygnus X-3 in both surveys. 
Therefore, this adds even more confidence
to its reality. In the left panel of Fig. \ref{surveys} we display
the CGPS image. These traces of extended emission are better seen here than in the GB6 single dish survey with lower angular resolution.

The presence of this extended emission around Cygnus X-3 in both surveys 
encouraged us to attempt a crude estimate of its spectral index. 
Our VLA (6 cm) and Ryle Telescope (2 cm) images
are likely missing some flux density from very short baselines and we preferred not to use them for such purpose. 
Since we were only interested in extended emission, the central core of Cygnus X-3 was initially
subtracted from both the CGPS and GB6 images. The CGPS image was later appropriately convolved
with an elliptical Gaussian so that its one arc-minute angular resolution matched the lower one of the GB6 (about 3 arc-minute).
In the process, we also transformed its brightness scale from brightness temperature to Jy beam$^{-1}$. 
The CGPS image was also regrided from galactic to equatorial coordinates.

Once the CGPS and GB6 images were correctly matched, we combined them in order to create a map of spectral index $\alpha$
($S_{\nu} \propto \nu^{\alpha}$).
The result is reported in the right panel of Fig. \ref{surveys}. Here, the spectral index of the extended radio feature under discussion
hints at a negative value $\alpha \simeq -0.5$ 
and suggesting a non-thermal emission mechanism. This would
be in natural agreement with relativistic electrons accelerated in the termination shock of a distorted
jet. In contrast, other bright extended
radio sources in the field have $\alpha$ values predominantly positive as expected from ordinary
star forming regions with thermal emission mechanisms. In any case, true matching beam observations at different
wavelengths are still needed to confidently constrain the spectral index value.

\section{Constraining the system energetics}

The possible Cygnus X-3 lobe or extended radio feature of other kind
has an integrated flux density of about 133 mJy at the 6 cm wavelength and extends
over a solid angle of about $6.8\times 10^{-7}$ sr. These values have been measured using the full multi-configuration
radio map in Fig. \ref{cd+d+b}.
For $\alpha=-0.5$, this implies a radio luminosity of $6.8 \times 10^{32}$ erg s$^{-1}$ in the 0.1-100 GHz at 9 kpc. 
Assuming standard
equipartition conditions, the minimum energy content is estimated as $3.2 \times 10^{47}$ erg with the equipartition magnetic
field being $2\times 10^{-5}$ G. 

The radio lobe dynamics in a microquasar environment using a simple analytic model has been studied by \cite{h2002}.
The typical size $r_e$, when the lobe reaches pressure equilibrium with the ISM, is given by:
\begin{equation}
r_e \sim 0.2~{\rm pc}~E_{44}^{1/3} p_{-11}^{-1/3},  \label{req}
\end{equation}
where $E_{44}$ represents the accumulated energy from multiple ejection events in units
of $10^{44}$ erg and $p_{-11}$ the external ISM pressure in units of $10^{-11}$ erg cm$^{-3}$.  
This equilibrium size is reached in a time scale of order:
\begin{equation}
t_e \sim 4\times 10^3~{\rm yr}~ E_{44}^{1/3} n_x^{1/2} p_{-11}^{-5/6} \label{teq}
\end{equation}
being $n_x$ the corresponding external ISM number density in units of cm$^{-3}$.

Assuming typical properties for intercloud gas in the Galaxy, such as $n_x\simeq 0.3$ cm$^{-3}$ and a $\sim 6000$\,K temperature,
a conceivable ISM pressure $p_{-11}=0.025$ is obtained. From the equipartition estimates above, we can adopt
$E_{44} = 3200$ for the total lobe energy content. Incidently, 
$\sim10^{44}$ erg is roughly the total energy involved in a single
major radio outburst of Cygnus X-3 (\cite{m1992}). Based on Eqs. \ref{req} and \ref{teq}, the expected lobe size is
$r_e \sim 10$ pc and the lobe equilibrium time $t_e \sim 7 \times 10^5$ yr. The resulting lobe dimension is 
within a factor of two from the estimated $\sim 5$ pc size, if our extended radio source is at the same distance as Cygnus X-3. 
The estimated lobe age would also be consistently shorter or comparable with 
the expected life time of the WR star ($\sim 10^6$ yr) acting as mass donor and powering the
outbursts of the system.

\section{Conclusions}

\begin{enumerate}

\item We revisited the issue of large scale radio features associated to the microquasar Cygnus X-3.
By combining different archive VLA observations, we have been able 
to create a very deep (rms noise 9.5 $\mu$Jy beam$^{-1}$) radio
map with sensitivity to arc-minute angular scales. Cygnus X-3 appears superposed onto a 
diffuse radio emission with apparent non-thermal index with an angular size of a few arc-minutes extending
South and South-West from it. The reality of such feature is independently confirmed by Ryle Telescope observations and
when inspecting in detail maps from the CGPS and other survey data. 

\item With all cautions in mind, we tentatively suggest the possibility that such an extended emission could be
physically associated to Cygnus X-3. It could be either
a distorted lobe, plume or partial shell-like structure resulting from the accumulated flaring history of the microquasar
interacting with the ISM. If connected with Cygnus X-3, the feature
linear size would be $\sim5$ pc. This is of the same order as extended lobes or bubbles seen
around other microquasars. Based on simple equipartition assumptions, the observed size is
roughly in agreement with expectations from simple theory of radio lobe dynamics.

\item A triple radio source with FR II morphology has been also discovered at few
arc-minute from Cygnus X-3 in our deep VLA map. In addition to provide a curious `family picture' of two accreting sources
in the same shot, the angular proximity of the FR II compact core could
enable future proper motion studies of this microquasar using future generations of highly sensitive interferometers.

\end{enumerate}

\begin{acknowledgements}
{\small The authors acknowledge support by
grants AYA2004-07171-C02-02 and AYA2004-07171-C02-01 from the Spanish government, and FEDER funds.
This has been also supported by Plan Andaluz de Investigaci\'on
of Junta de Andaluc\'{\i}a as research group FQM322.
The NRAO is a facility of the NSF
operated under cooperative agreement by Associated Universities, Inc.
The Ryle Telescope is supported by the
UK Science and Technology Facilities Council.
This paper is also based on observations collected at the Centro Astron\'omico Hispano Alem\'an
(CAHA) at Calar Alto, operated jointly by the Max-Planck Institut
f\"ur Astronomie and the Instituto de Astrof\'{\i}sica de Andaluc\'{\i}a (CSIC).
J.A.C. is a researcher of the programme {\em Ram\'on y
Cajal} funded jointly by the Spanish Ministerio de Educaci\'on y Ciencia (former
Ministerio de Ciencia y Tecnolog\'{\i}a) and Universidad de Ja\'en.
This research made use of the SIMBAD
database, operated at the CDS, Strasbourg, France.
This research used the facilities of the Canadian Astronomy Data Centre operated by the National Research Council of Canada with the support of the Canadian Space Agency. 
The research presented in this paper has used data from the Canadian Galactic Plane Survey a Canadian project with international partners supported by the Natural Sciences and Engineering Research Council.
}
\end{acknowledgements}


\end{document}